\documentclass[aps,prb,twocolumn,superscriptaddress,showpacs]{revtex4}
\pdfoutput=1
\usepackage{amsmath}
\usepackage{amssymb}
\usepackage{url}
\usepackage{graphicx}% Include figure files
\usepackage{dcolumn} % Align table columns on decimal point
\usepackage{bm}      % bold math
\usepackage{multirow}
\usepackage{epstopdf}
\usepackage{color}
\usepackage{gensymb}
\usepackage{textcomp}
\usepackage{hyperref}

\begin{document}

\title{Evolution of the magnetic and structural properties of Fe$_{1-x}$Co$_x$V$_2$O$_{4}$}

\author{R.~Sinclair}
\affiliation{Department of Physics and Astronomy, University of Tennessee,
Knoxville, Tennessee 37996-1200, USA}

\author{J.~Ma}
\affiliation{Department of Physics and Astronomy, University of Tennessee,
Knoxville, Tennessee 37996-1200, USA}

\author{H.B.~Cao}
\affiliation{Quantum Condensed Matter Division, Oak Ridge National Laboratory, Oak Ridge, Tennessee 37381, USA}

\author{T.~Hong}
\affiliation{Quantum Condensed Matter Division, Oak Ridge National Laboratory, Oak Ridge, Tennessee 37381, USA}

\author{M.~Matsuda}
\affiliation{Quantum Condensed Matter Division, Oak Ridge National Laboratory, Oak Ridge, Tennessee 37381, USA}

\author{Z.~L.~Dun}
\affiliation{Department of Physics and Astronomy, University of Tennessee,
Knoxville, Tennessee 37996-1200, USA}

\author{H.~D.~Zhou}
\affiliation{Department of Physics and Astronomy, University of Tennessee, Knoxville, Tennessee 37996-1200, USA}
\affiliation{National High Magnetic Field Laboratory, Florida State University, Tallahassee, Florida 32310, USA}

\date{\today}% It is always \today, today,

\begin{abstract}
The magnetic and structural properties of single crystal Fe$_{1-x}$Co$_x$V$_2$O$_{4}$ samples have been investigated by performing specific heat, susceptibility, neutron diffraction, and X-ray diffraction measurements. As the orbital-active Fe$^{2+}$ ions with larger ionic size are gradually substituted by the orbital-inactive Co$^{2+}$ ions with smaller ionic size, the system approaches the itinerant electron limit with decreasing V-V distance. Then, various factors such as the Jahn-Teller distortion and the spin-orbital coupling of the Fe$^{2+}$ ions on the A sites and the orbital ordering and electronic itinerancy of the V$^{3+}$ ions on the B sites compete with each other to produce a complex magnetic and structural phase diagram. This phase diagram is compared to those of  Fe$_{1-x}$Mn$_x$V$_2$O$_{4}$ and Mn$_{1-x}$Co$_x$V$_2$O$_{4}$ to emphasize several distinct features.
\end{abstract}


\maketitle

\subsection{I. INTRODUCTION}
Normal spinels\cite{AV2O4review} AV$_2$O$_4$ (A = Cd, Mn, Fe, Mg, Zn, and Co) have received considerable attention due to their physical properties resulting from the interplay among the spin-lattice coupling from the localized 3$d$ electrons, the orbital degrees of freedom, and the geometrically frustrated structure. Furthermore, the AV$_2$O$_4$ normal spinels can be divided into two groups based on the A ions. One group includes AV$_2$O$_4$ (A = Cd\cite{Cd1, Cd2, Cd3}, Mg\cite{Mg1, Mg2, Mg3, Mg4, Mg5}, Zn\cite{Zn1, Zn2}) with non-magnetic A ions. In these three materials, the orbital ordering (OO) transition drives a cubic to tetragonal structural phase transition at low temperatures which relieves the geometrical frustration of the V-pyrochlore sublattice and leads to an antiferromagnetic transition of the V$^{3+}$ ions. 
 
The other group includes AV$_2$O$_4$ (A = Mn, Fe, Co) with magnetic A ions in which the additional A-B magnetic interactions or the Jahn-Teller (JT) active Fe$^{2+}$ ions lead to more complex physical properties. For example, (i) MnV$_2$O$_4$\cite{Mn1, Mn2, Mn3, Mn4, Mn5, Mn6, Mn7} exhibits a magnetic phase transition at 56 K with a collinear ferrimagnetic (CF) structure where the Mn$^{2+}$ moments are antiparallel to the V$^{3+}$ moments. Then, an antiferro-OO transition in the $t_{2g}$ orbitals of the V$^{3+}$ ions occurs at 53 K, where the $d_{xy}$ orbital is occupied by one electron and the other electron occupies the $d_{yz}$ and $d_{zx}$ orbitals alternately along the $c$-axis. The characteristic feature of this OO transition is the accompanied cubic to tetragonal structural transition involving a compressed tetragonal distortion ($c$ $<$ $a$). This OO transition also results in a non-collinear ferrimagnetic (NCF) ordering below 53 K where the V$^{3+}$ moments are canted from the $[$111$]$ direction; (ii) CoV$_2$O$_4$\cite{Co1, Co2, Co3} exhibits two magnetic transitions at 150 K and 75 K which are CF and NCF transitions, respectively. This sample also shows no OO transition due to the fact that it is approaching the itinerant electron behavior with the small V-V distance. In the AV$_2$O$_4$ system, this increased electronic itinerancy due to the decreased V-V distance has been theoretically predicted\cite{Goodenough, Fran2} and experimentally confirmed\cite{Rogers1, Rogers2, Fran1}; (iii) FeV$_2$O$_4$\citep{Fe1, Fe2, Fe3, Fe4, Fe5, Fe6} exhibits at least three transitions. It is unique since the Fe$^{2+}$ (3$d^6$) ions have orbital degrees of freedom in the doubly degenerate $e_g$ states. First, a structural transition from a cubic to a tetragonal phase ($c$ $<$ $a$) occurs at 140 K which mainly involves the OO transition of Fe$^{2+}$ ions. Then, a second structural transition from a tetragonal to an orthorhombic phase occurs    at 110 K which is accompanied by a CF transition. Finally, a third structural transition from an orthorhombic to another tetragonal phase ($c$ $>$ $a$) occurs at 60 K which is accompanied with a NCF transition. In this low temperature tetragonal phase with $c$ $>$ $a$ , a  ferro-OO transition containing a complex orbital of the V$^{3+}$ ions has been proposed\citep{Fe5} which is in contrast to the OO of the real V-orbitals observed in the tetragonal phase with $c$ $<$ $a$ for MnV$_2$O$_4$. 

To better understand the distinct physical properties among AV$_2$O$_4$ (A = Mn, Fe, Co), several studies on these solid solutions have been conducted. For example, the resistivity and X-ray diffraction (XRD) studies on Mn$_{1-x}$Co$_x$V$_2$O$_{4}$ show that with increasing Co-doping, the system approaches the itinerant electron limit with decreasing resistivity\cite{MnCo1}. Around $x$ = 0.8, the system shows no structural phase transition down to 10 K\cite{MnCo1, MnCo2}. Recently, the neutron scattering experiments and first principle calculations have revealed that the strong competition between the orbital ordering and itinerancy in Mn$_{1-x}$Co$_x$V$_2$O$_{4}$ is the key factor for its complex magnetic and structural phase diagram. Interestingly, both the orbital ordering in the low Co-doping samples and the magnetic isotropy in the high Co-doping samples lead to the NCF states\cite{MnCo3}. Modern studies on Fe$_{1-x}$Mn$_x$V$_2$O$_{4}$\cite{FeMn1, FeMn2} also reveal a complex phase diagram in which the ferro-OO is gradually suppressed with increasing $x$ and changes to the antiferro-OO for $x$ $>$ 0.6. Around $x$ = 0.6,  the long range orbital ordering of the Fe $^{2+}$ ions also disappears. This indicates that the ferro-OO is possibly stabilized by the orbital degrees of freedom of the Fe$^{2+}$ ions located at the A site. 

In this paper, we aim to study the magnetic and structural properties of another solid solution of V-spinels:  Fe$_{1-x}$Co$_x$V$_2$O$_{4}$. The detailed specific heat, susceptibility, neutron diffraction, and XRD measurements performed on single crystals of Fe$_{1-x}$Co$_x$V$_2$O$_{4}$ reveal a complex magnetic and structural phase diagram.

\subsection{II. EXPERIMENTAL METHODS}
Single crystals of Fe$_{1-x}$Co$_x$V$_{2}$O$_{4}$ were grown by the traveling-solvent floating-zone (TSFZ) technique. The feed and seed rods for the crystal growth were prepared by solid state reactions. Appropriate mixtures of FeO, CoO and V$_{2}$O$_{3}$ were ground together and pressed into 6-mm-diameter 60-mm rods under 400 atm hydrostatic pressure and then calcined in vacuum in a sealed quartz tube at 950 $^{\circ}$C for 12 hours.  The crystal growth was carried out in argon in an IR-heated image furnace (NEC) equipped with two halogen lamps and double ellipsoidal mirrors with feed and seed rods rotating in opposite directions at 25 rpm during crystal growth at a rate of 15 mm/h. Small pieces of single crystals were ground into fine, flat-plate powder samples for XRD, and the diffraction patterns were recorded with a HUBER Imaging Plate Guinier Camera 670 with Ge monochromatized Cu $K_{\alpha1}$ radiation (1.54059 {\AA}). Data was collected at temperatures down to 10 K with a cryogenic helium compressor unit. The lattice parameters were refined from the XRD patterns by using the program $FullProf$ with typical refinements for all samples having $\chi^2$ $\approx$ 0.3. The refinements also corrected for the absorbed radiation. X-ray Laue diffraction was used to align the crystals. The dc magnetic-susceptibility measurements were performed using a Quantum Design superconducting interference device (SQUID) magnetometer using a magnetic field of 0.01 T. The specific heat measurements were performed on a Quantum Design Physical Property Measurement System (PPMS). The crystal samples used for the magnetic-susceptibility and the specific heat measurements were not aligned. Neutron-diffraction experiments were performed at the four-circle diffractometer (HB-3A) and the cold neutron triple-axis spectrometer (CG-4C, CTAX) configured to measure along the (H,H,K) planes at the High Flux Isotope Reactor (HFIR) of the Oak Ridge National Laboratory (ORNL).

\subsection{III. EXPERIMENTAL RESULTS}

\subsection{A. phase diagram}

\begin{figure}[tbp]
\linespread{1}
\par
\begin{center}
\includegraphics[width=3.6in]{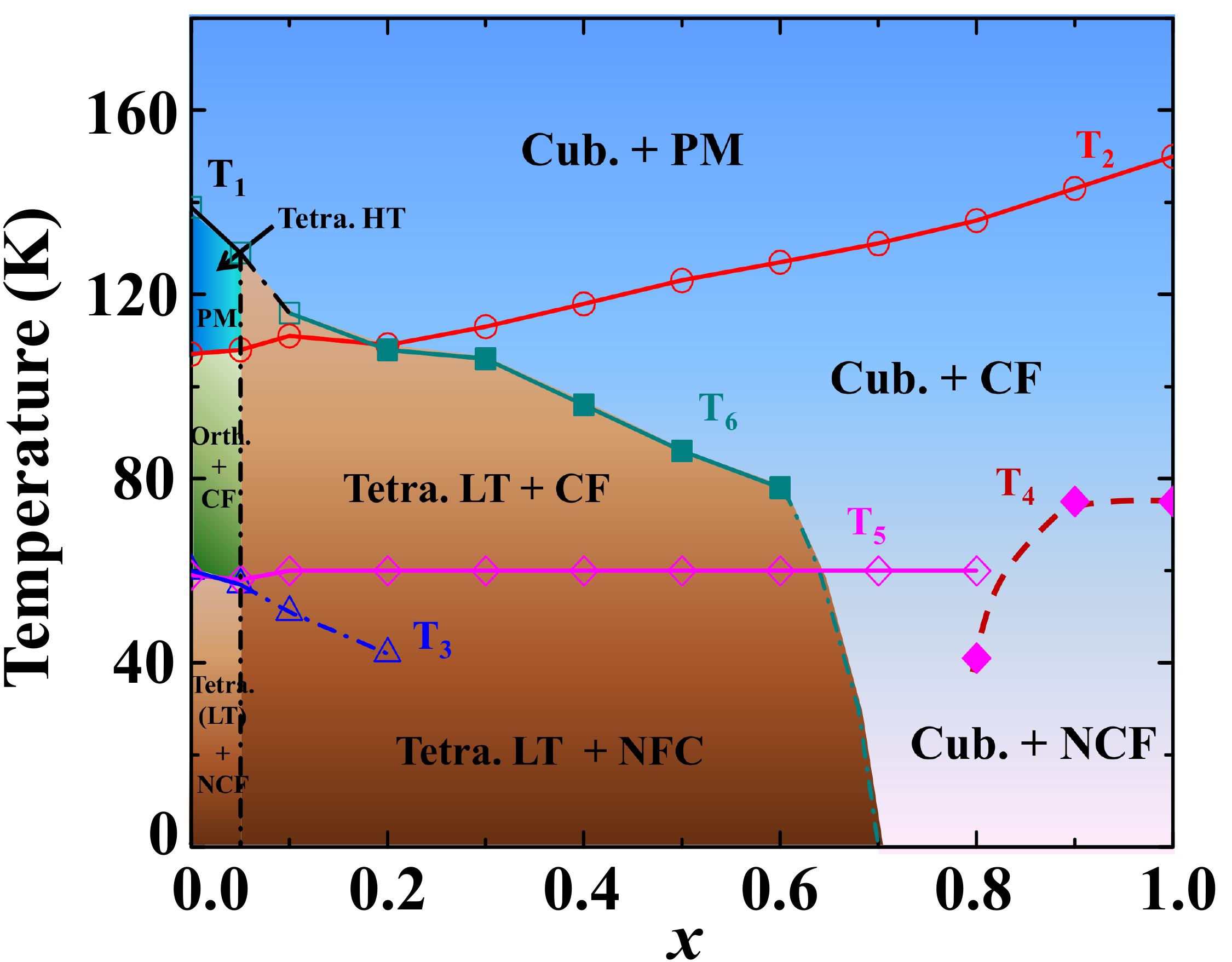}
\end{center}
\par
\caption{(Color Online) The magnetic and structural phase diagram of Fe$_{1-x}$Co$_x$V$_2$O$_{4}$. $T_1$ ($\Box$) represents the cubic to high temperature (HT) tetragonal ($c$ $<$ $a$) phase transition for the $x$ = 0.0 and 0.05 samples and the cubic to low temperature (LT) tetragonal ($c$ $>$ $a$) phase transition for the $x$ = 0.1 sample; $T_2$ ($\bigcirc$) represents the paramagnetic to CF transition; moreover, it represents the HT tetragonal to orthorhombic phase transition for the $x$ = 0.0 and 0.05 samples; $T_3$ ($\vartriangle$) represents the orthorhombic to LT tetragonal phase transition for the $x$ = 0.0 and 0.05 samples and the structural distortion within the LT tetragonal phase for the $x$ = 0.1 and 0.2 samples; $T_4$ ($\blacklozenge$) represents the CF to NCF transition for the $x$ $\geq$ 0.8 samples; $T_5$ ($\lozenge$) represents the CF to NCF transition for the $x$ $\leq$ 0.8 samples; $T_6$ ($\blacksquare$) represents the cubic to LT tetragonal phase transition for the  0.2 $\leq$ $x$ $\leq$ 0.6 samples. }
\end{figure}

Based on the following structural and magnetic data, a phase diagram of Fe$_{1-x}$Co$_x$V$_2$O$_{4}$ is constructed, as shown in Fig. 1. Due to the large number of transitions, the phase diagram is presented first to introduce the general trends observed in the data. In total, our samples exhibited six different transitions which were corroborated through several experimental techniques. The description of each transition is detailed in the caption of Figure 1. 

\subsection{B. susceptibility and specific heat}

\begin{figure*}[tbp]
\linespread{1}
\par
\begin{center}
\includegraphics[width=5.2in]{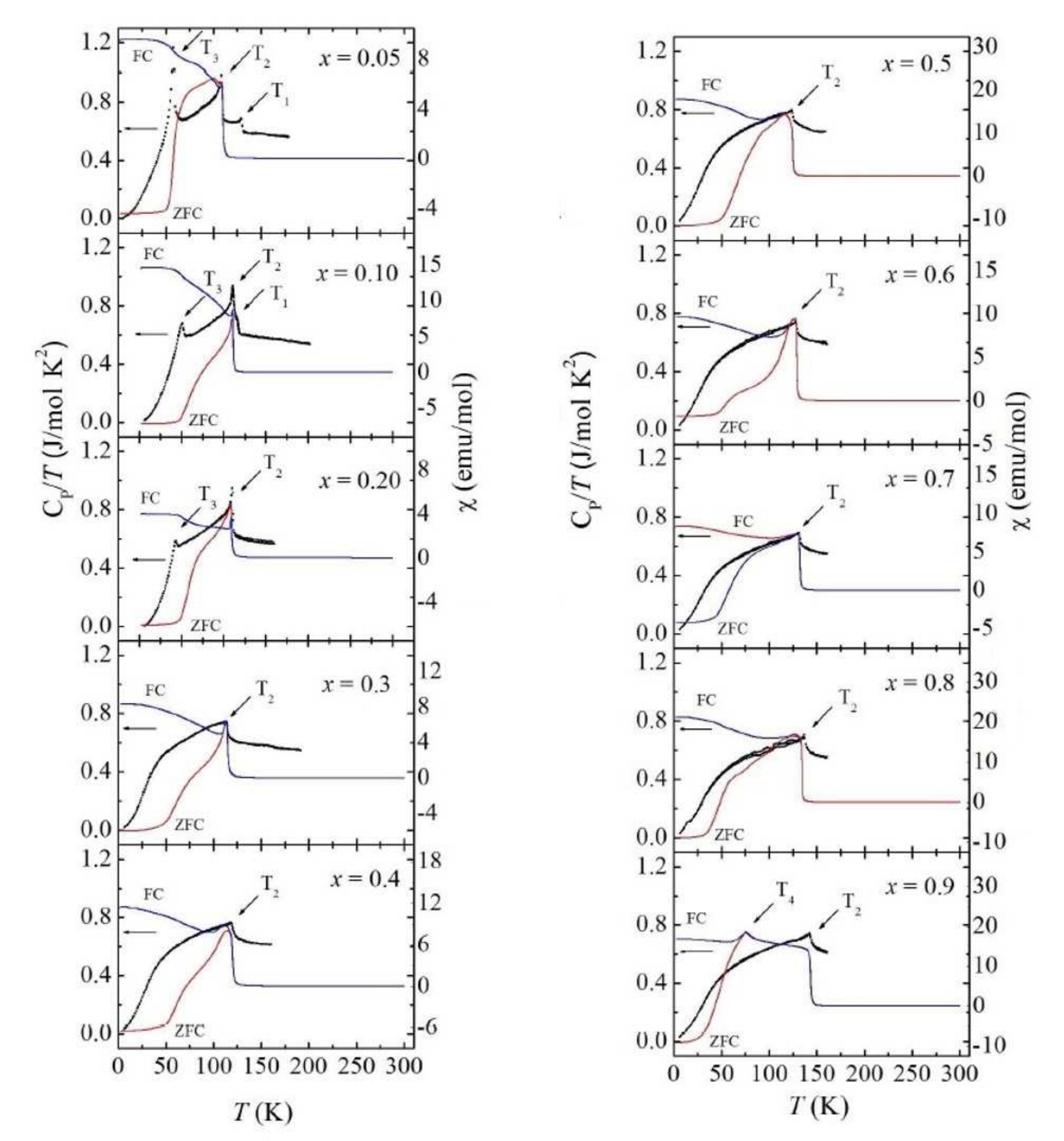}
\end{center}
\par
\caption{(Color Online) The temperature dependencies of the dc magnetic susceptibility and specific heat for Fe$_{1-x}$Co$_x$V$_2$O$_{4}$.}
\end{figure*}

Figure 2 shows the temperature dependence of the dc magnetic susceptibility and specific heat for Fe$_{1-x}$Co$_x$V$_2$O$_{4}$. For $x$ = 0.05, the specific heat shows three transitions at $T_1$ = 129 K, $T_2$ = 108 K, and $T_3$ = 57 K. At $T_2$, the susceptibility shows a sharp increase. At $T_3$, the zero field cooling susceptibility (ZFC) shows a sharp drop. By comparing these transitions to FeV$_2$O$_4$ ($T_1$ = 139 K, $T_2$ = 109 K, and  $T_3$ = 60 K), it is obvious that  $T_1$ represents the cubic to high temperature (HT) tetragonal ($c$ $<$ $a$) phase transition,  $T_2$ represents the CF transition with the tetragonal to orthorhombic phase transition, and  $T_3$ represents the NCF  transition with the orthorhombic to low temperature (LT) tetragonal ($c$ $>$ $a$) phase transition. It is also obvious that with 5\% Co doping, both  $T_1$ and  $T_3$ decrease but  $T_2$ increases. For $x$ = 0.1, the specific heat still shows three transitions. The susceptibility also still shows a sharp increase at  $T_2$, but the ZFC susceptibility does not show a sharp decrease at  $T_3$ any more. For $x$ = 0.2, the specific heat just shows two peaks at  $T_2$ and  $T_3$. Since at the first peak temperature the susceptibility shows a sharp increase, we assigned this as  $T_2$. For 0.3 $\leq$ $x$ $\leq$ 0.8, the specific heat shows only one peak at  $T_2$, where again the related susceptibility shows a sharp increase. For $x$ = 0.9, the susceptibility shows a  sharp increase at $T_2$ and another clear cusp around  $T_4$ = 75 K. By comparing these transitions to CoV$_2$O$_4$, one would expect that  $T_2$ and $T_4$ correspond to the CF and NCF ordering temperatures, respectively.

\begin{figure}[tbp]
\linespread{1}
\par
\begin{center}
\includegraphics[width=3.6 in]{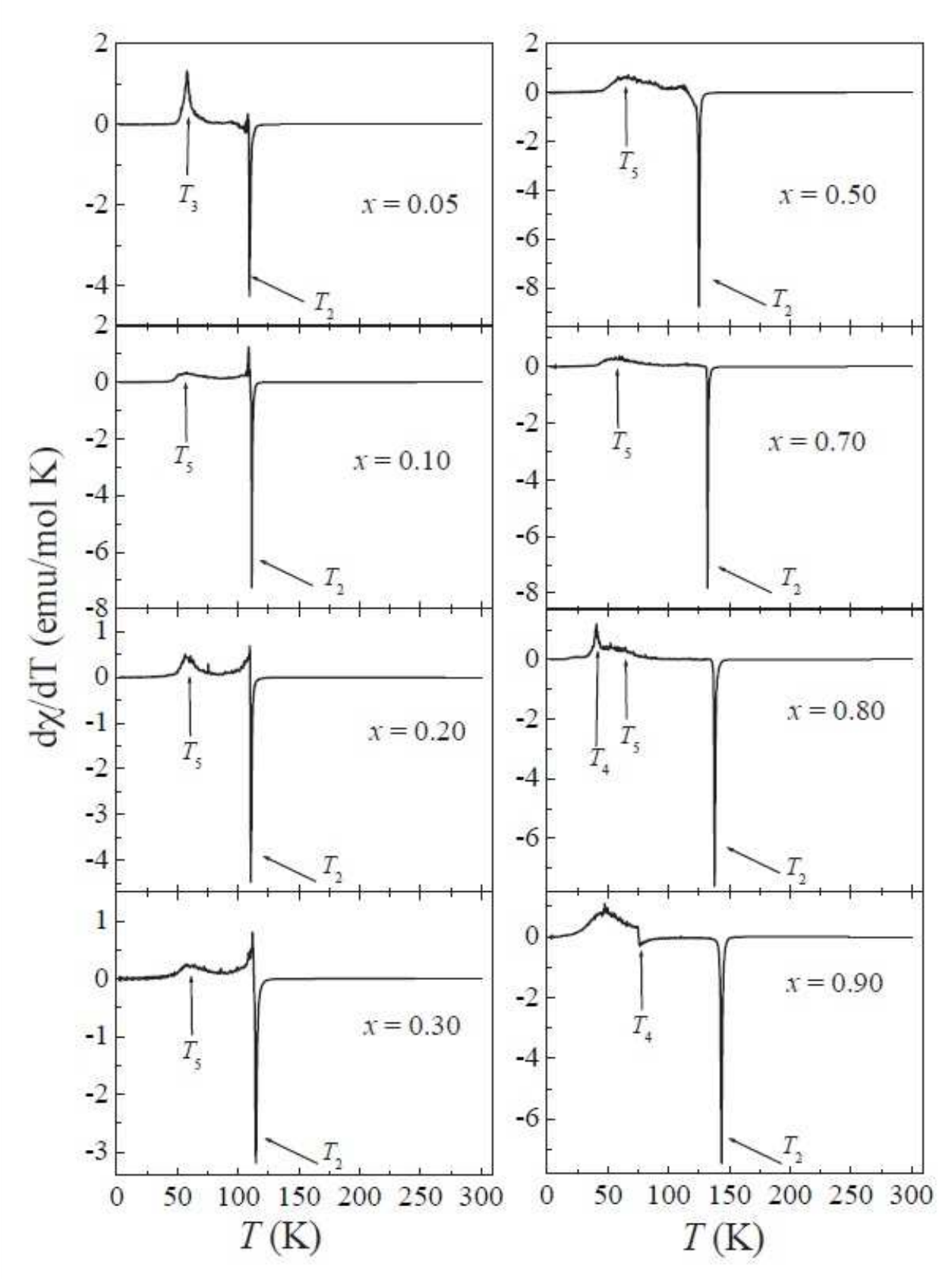}
\end{center}
\par
\caption{The derivative of the ZFC susceptibility for Fe$_{1-x}$Co$_x$V$_2$O$_4$.}
\end{figure}

In order to probe the magnetic phase transition of Fe$_{1-x}$Co$_x$V$_2$O$_{4}$ in more detail, the derivative of the ZFC susceptibility is shown in Fig. 3. For $x$ = 0.05, the derivative shows two sharp peaks at  $T_2$ and  $T_3$. For 0.1 $\leq$ $x$ $\leq$ 0.7, every sample's derivative shows a broad peak around 60 K as well as the sharp peak at $T_2$. It is noteworthy that this 60 K (we assigned this temperature as $T_5$) feature is not exactly at the $T_3$ temperatures for $x$ = 0.1 ($T_3$ = 51 K) and 0.2 ($T_3$ = 42 K) samples observed from the specific heat. For $x$ = 0.8, below the broad peak at 60 K, there is another sharp peak around 40 K. For $x$ = 0.9, the derivative shows  a sharp peak at $T_2$ and a jump at $T_4$.

The specific heat and susceptibility show complex magnetic and structural evolution for Fe$_{1-x}$Co$_x$V$_2$O$_{4}$. Several general trends are that with increasing Co-doping ($x$), (i) $T_1$ decreases and disappears with $x$ $\geq $ 0.2; (ii) $T_2$ increases; (iii) $T_3$ decreases and disappears with $x$ $\geq$ 0.3; (iv) $T_5$ ($\sim$ 60 K) seems to be Co-doping independent for 0.1 $\leq$ $x$ $\leq$ 0.7.

\subsection{C. single crystal neutron diffraction}

\begin{figure}[tbp]
\linespread{1}
\par
\begin{center}
\includegraphics[width=3.0 in]{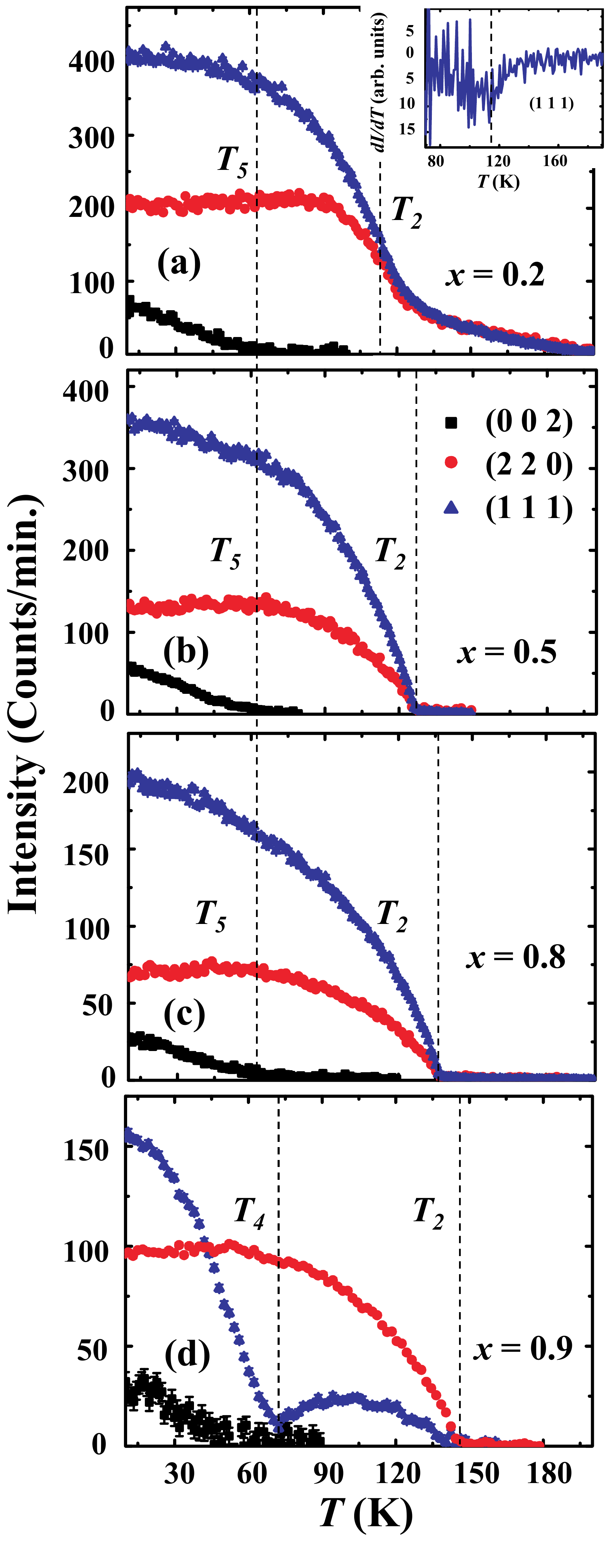}
\end{center}
\par
\caption{ (Color online) The temperature dependence of the Bragg peak intensities, (002) (squares), (220) (circles), and (111) (triangles) for Fe$_{1-x}$Co$_x$V$_2$O$_{4}$ measured on HB-3A, (a) $x$ = 0.2, (b)$x$ = 0.5, (c) $x$ = 0.8, and (d) $x$ = 0.9. Insert of (a): The derivative of the intensity with respect to the temperature of the (111) Bragg peak.}
\end{figure}

To further clarify the magnetic phase transitions in Fe$_{1-x}$Co$_x$V$_2$O$_{4}$, single crystal neutron diffraction measurements have been performed on selective samples. Figure 4 shows the temperature dependence of the intensity of several Bragg peaks ((002), (220), (111)) of these samples. With increasing Co-doping, both the magnetic moments and the V-canting angles decrease compared with FeV$_2$O$_4$; however, the structural transition that the $x$ = 0.2 and $x$ = 0.5 samples undergo make it difficult to determine the exact values of the moments and canting angles from single crystal neutron diffraction. For the $x$ = 0.8 sample at 5 K, the total moment of the A site ions is 3.2(1) $\mu_{B}$ and the total moment of the B site (V$^{3+}$) ions is 0.8(2) $\mu_{B}$, and the V-canting angle decreases from 55(4)$\degree$ for FeV$_2$O$_{4}$to 38(3)$\degree$\cite{Fe4}. For $x$ = 0.2,  a ferrimagnetic (FIM) signal develops below 109 K ($T_2$) at the symmetry-allowed Bragg positions (220) and (111)  which confirms the paramagnetic to CF transition. While the (002) peak is forbidden by the symmetry, the observed scattering intensity below 60 K signals the formation of an antiferromagnetic (AFM) spin structure in the $ab$ plane. Therefore, the onset of the (002) magnetic reflection marks the CF-NCF transition at $T_5$. For the $x$ = 0.5 and $x$ = 0.8 samples, the onset of (002) peak occurs around 60 K ($T_5$) as well. Similar behaviors of (220) and  (111) peaks of the $x$ = 0.5, 0.8, and 0.9 samples confirm the CF transition at $T_2$. For $x$ = 0.9, the (002) peak behavior also confirms its CF-NCF transition at 75 K ($T_4$). Also note that for the $x$ = 0.1 and 0.2 samples, $T_3$ no longer represents the CF-NCF transition since at higher temperatures, $T_5$, the NCF ordering already occurs.

\subsection{D. X-ray diffraction}

\begin{figure}[tbp]
\linespread{1}
\par
\begin{center}
\includegraphics[width=3.6in]{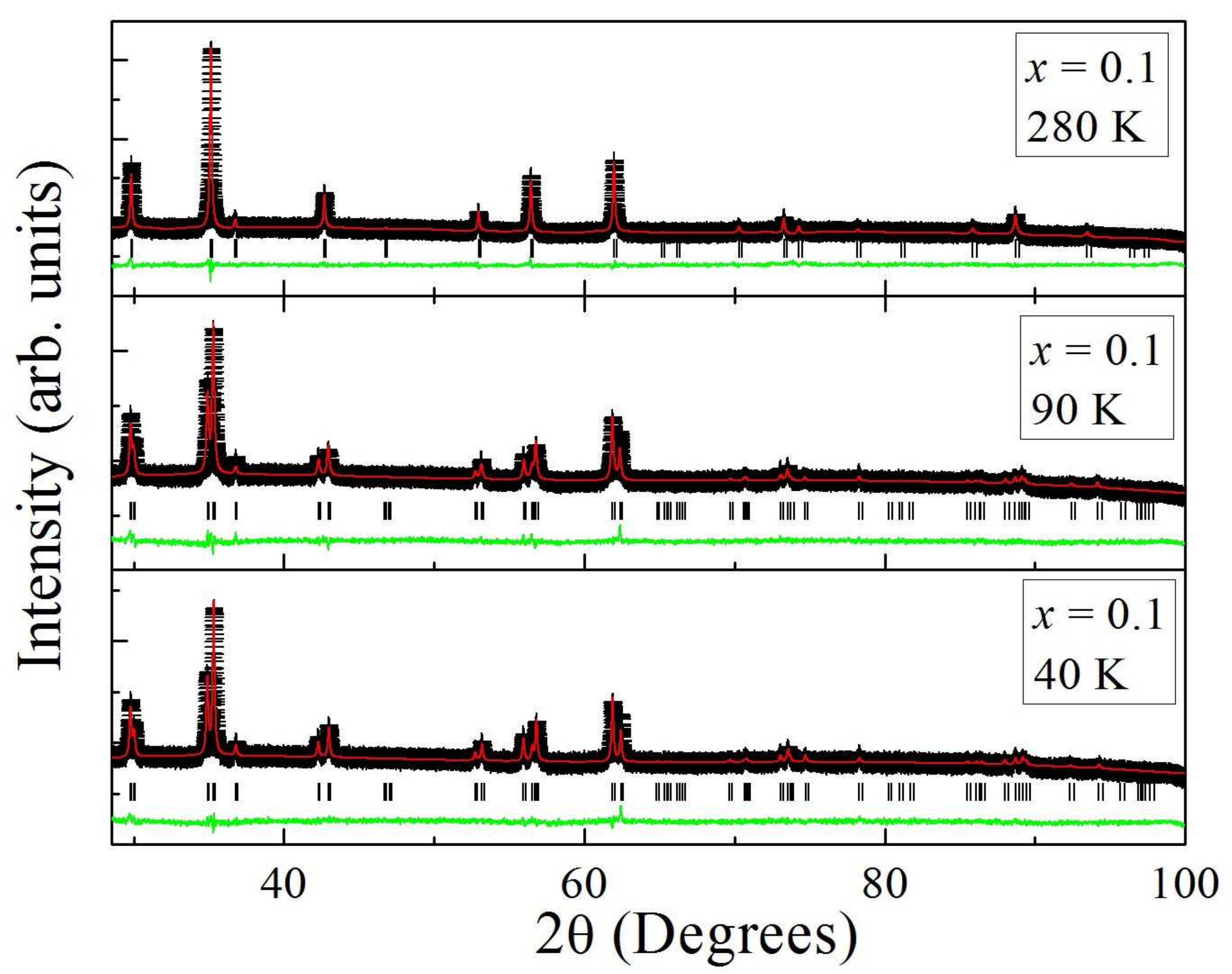}
\end{center}
\par
\caption{(Color Online) The XRD patterns for $x$ = 0.1 sample (squares)
at 280 K (a), 90 K (b), and 40 K (c). The solid curves are the best fits from
the Rietveld refinement using $FullProf$. The vertical marks indicate
the position of Bragg peaks, and the bottom curves show the
difference between the observed and calculated intensities. }
\end{figure}

\begin{figure}[tbp]
\linespread{1}
\par
\begin{center}
\includegraphics[width=3.4in]{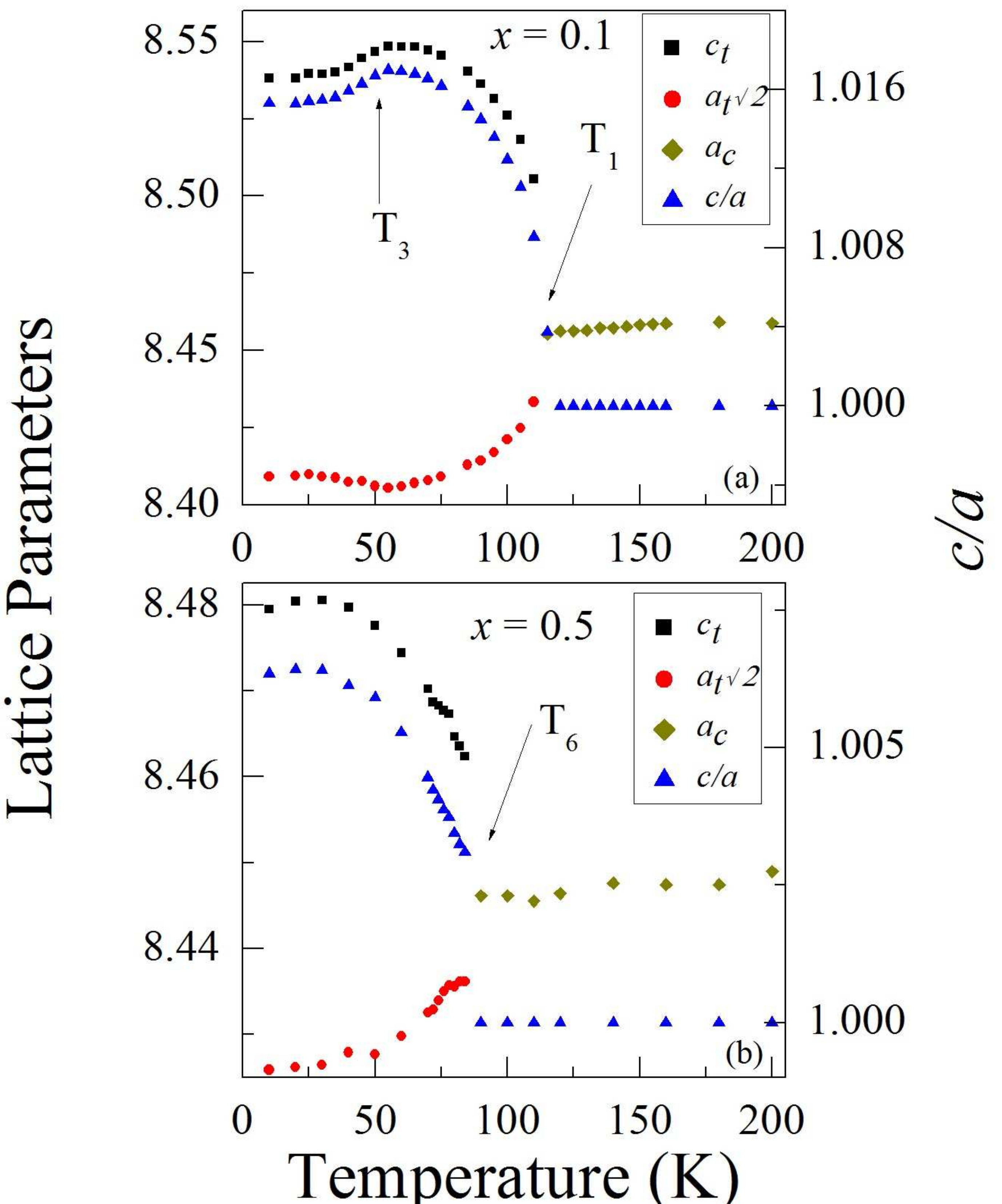}
\end{center}
\par
\caption{(color online) (a) The temperature dependence of the lattice parameters and $c/a$ ratio for $x$ = 0.1 sample. (b) The temperature dependence of the lattice parameters and $c/a$ ratio for the $x$ = 0.5 sample. }
\end{figure}

\begin{figure}[tbp]
\linespread{1}
\par
\begin{center}
\includegraphics[width=3.6in]{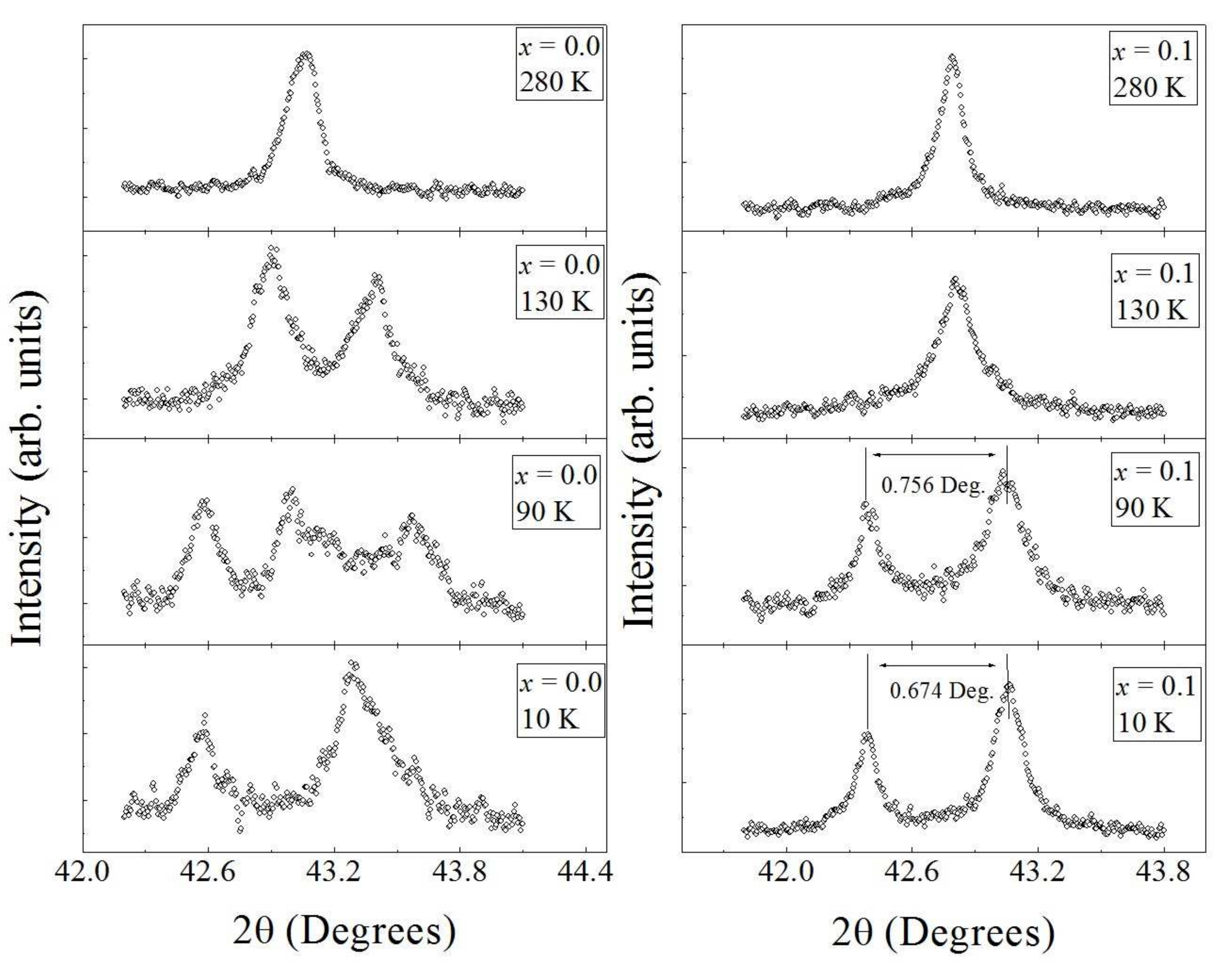}
\end{center}
\par
\caption{The temperature dependence of the (400) peaks for FeV$_2$O$_4$ and  $x$ = 0.1 samples at different temperatures. }
\end{figure}

To better understand the structural phase transition in Fe$_{1-x}$Co$_x$V$_2$O$_{4}$, XRD measurements down to 10 K were performed. Figure 5 shows the measured patterns and related refinements for $x$ = 0.1 at 280 K, 90 K, and 40 K, respectively. At high temperature (280 K), the sample has a cubic phase. At 90 K $<$ $T_1$ = 111 K, the best refinement of the pattern leads to a tetragonal structure ($I4$$_1$$/amd$) with $c$ $>$ $a$. Then at 40 K $ < $ $T_3$ = 51 K, the refinement shows that it keeps the same tetragonal structure. Here, we tested the XRD pattern at 40 K with all the three possible tetragonal phases reported for FeV$_2$O$_4$ (the HT tetragonal phase($I4$$_1$$/amd$) with $c$ $<$ $a$ and the LT tetragonal phase ($I4$$_1$$/amd$) with $c$ $>$ $a$) and MnV$_2$O$_4$ (the tetragonal phase ($I4$$_1$$/a$) with $c$ $<$ $a$). The major difference among these three phases are the atomic positions for Fe(Mn) and V ions\cite{Fe4, Fe5}. The refinements using the three phases lead to consistent results with $c$ $>$ $a$, and the tetragonal phase ($I4$$_1$$/amd$) with $c$ $>$ $a$ gives the best fitting results. The temperature dependence of the lattice parameters for $x$ = 0.1 obtained from the detailed XRD measurements is shown in Fig. 6(a). Around $T_1$ = 111 K, the cubic phase changes to the tetragonal phase with $c$ $>$ $a$. Then below $T_3$ = 51 K, the lattice parameter $c$ slightly decreases and $a$ slightly increases which leads to a decrease of the $c/a$ ratio . 

It is obvious that the structural transitions for $x$ = 0.1 are different from those of FeV$_2$O$_4$. To further demonstrate this difference, the temperature dependence of the (400) peak for both samples are shown in Fig. 7. For FeV$_2$O$_4$, the single (400) peak splits to two peaks below $T_1$ = 139 K (cubic to HT tetragonal phase), then splits to three peaks below $T_2$ = 107 K (HT tetragonal to orthorhombic phase), then merges to two peaks again (orthorhombic to LT tetragonal phase) below $T_3$ = 60 K. But for $x$ = 0.1, the single (400) peak splits to two peaks ((400) and (004)) just below $T_1$  = 111 K. With decreasing temperature these two peaks move away from each other or the splitting 2$\theta$ within these two peaks increases, which means the $c/a$ ratio increases. Then below $T_3$ = 51 K, these two peaks begin to move towards each other, which means the $c/a$ ratio decreases. As shown in Fig. 7, the splitting for $x$ = 0.1 sample is 0.756 degree at 60 K but 0.674 degree at 10 K. This subtle structural distortion at $T_3$ occurs below its CF-NCF magnetic transition at $T_5$ = 60 K. While both the HT tetragonal and orthorhombic phases still manifest in the $x$ = 0.05 sample, in the $x$ = 0.1 sample the HT tetragonal ($c$ $<$ $a$) and orthorhombic phases do not exist. Its structure changes from the cubic to the LT tetragonal  ($c$ $>$ $a$) phase directly. The refinements of the XRD data for the $x$ = 0.2 sample show similar results as those of the $x$ = 0.1 sample (not shown here).

\begin{figure}[tbp]
\linespread{1}
\par
\begin{center}
\includegraphics[width=3.4in]{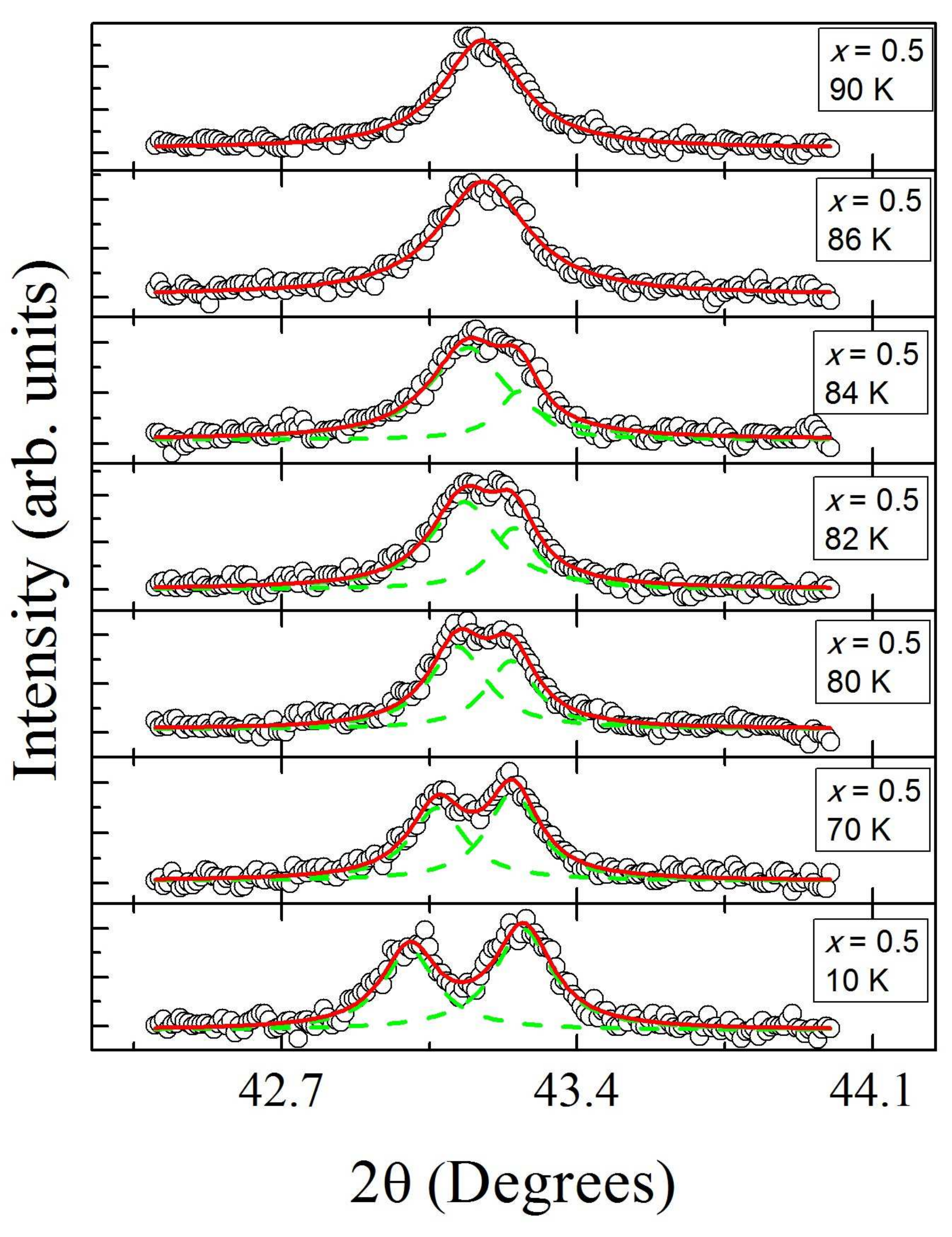}
\end{center}
\par
\caption{(color online) The temperature dependence of the (400) peak for the $x$ = 0.5 sample. The green dashed line represents the Lorentzian fit. The red solid line represents the total fitting. }
\end{figure}

The temperature dependence of the lattice parameters and $c/a$ ratio for $x$ = 0.5 (Fig. 6(b)) show that there is a cubic to tetragonal phase (($I4$$_1$$/amd$) with $c > a$) transition at 84 K. This temperature is below $T_2$ ( the CF ordering temperature) and above $T_5$ = 60 K (the NCF ordering temperature). We assigned this temperature as $T_6$. As shown in Fig. 8, the single (400) peak splits around 84 K, which confirms the structural phase transition at $T_6$. The XRD refinements for samples with 0.3 $\leq$ $x$  $\leq$0.6 show similar structural phase transition at 106 K for $x$ = 0.3, 96 K for $x$ = 0.4 and 78 K for $x$ = 0.6, respectively. The general trend is that with increasing $x$, $T_6$ decreases. There is no further structural phase transition or distortion below $T_6$ for 0.3 $\leq$ $x$  $\leq$0.6.  For the $x$ $\geq$ 0.7 samples, there is no structural phase transition down to 10 K according to the XRD data (not shown here). 

\begin{figure}[tbp]
\linespread{1}
\par
\begin{center}
\includegraphics[width=3.4in]{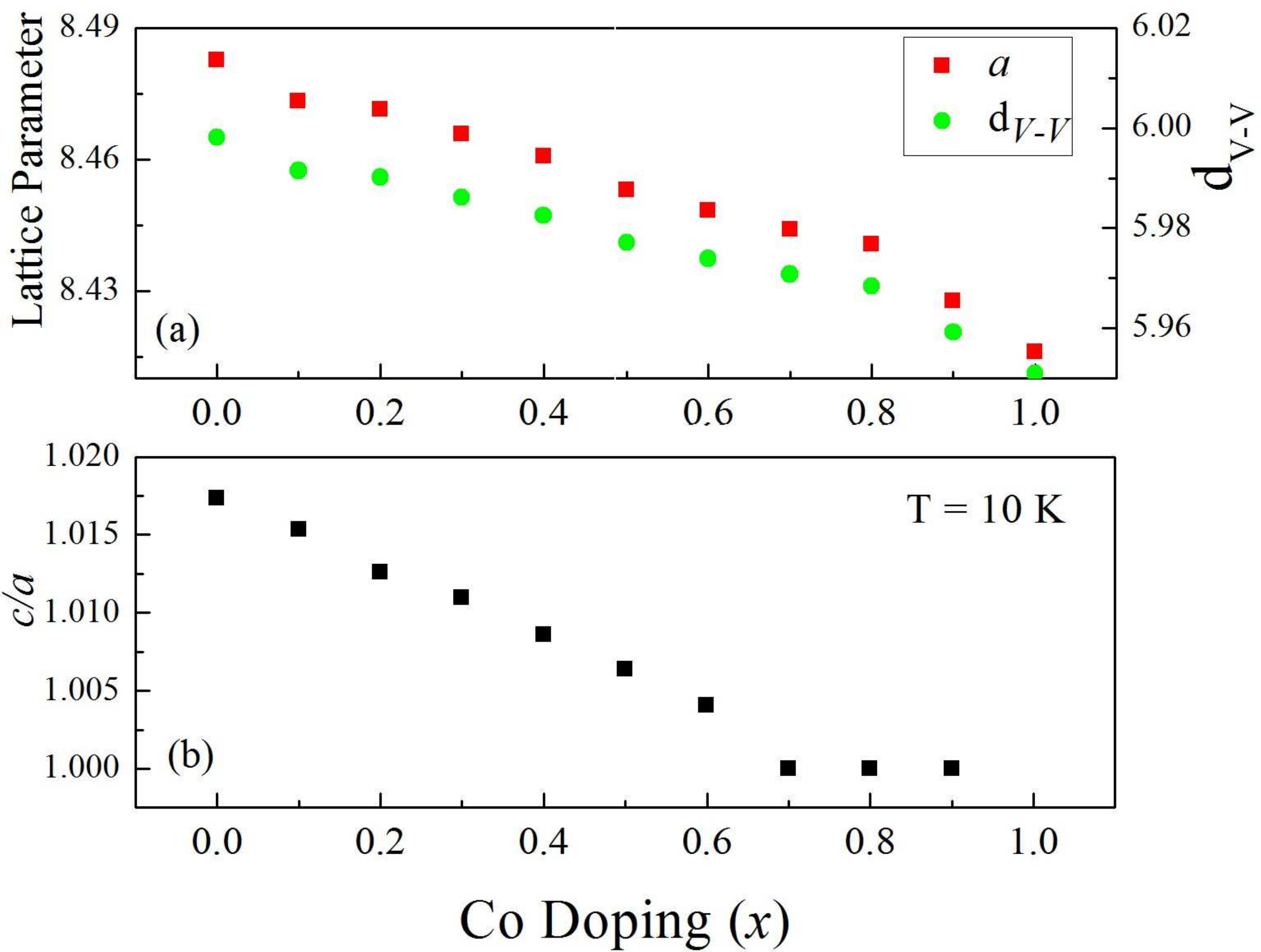}
\end{center}
\par
\caption{(color online) The Co-doping dependence of (a) the lattice parameter $a$ and $d_{V-V}$ at room temperature; and (b) the $c/a$ ratio at 10 K. }
\end{figure}

Another general rule obtained from the XRD refinements is that at room temperature the lattice parameter $a$ and the distance between the nearest V ions ($d_{V-V}$) decreases with increasing Co-doping, as shown in Fig. 9(a). At 10 K, the $c/a$ ratio increases (the distortion decreases) with increasing Co-doping. The structural parameters for the $x$ = 0.1 and 0.5 samples at room temperature and 10 K are listed in Table I.

\begin{table*}[tbp]
\par
\caption{Structural parameters for the $x$ = 0.1 and 0.5 samples at 280 K (space group $Fd-3m$) and 10 K (space group $I4$$_1$$/amd$). The B-values for the Oxygen atoms presented below were optimized to find the best fit, and the values are have larger uncertainty due to the relatively low energy and flux of the laboratory X-ray diffractometer used. }
\label{t1}
\setlength{\tabcolsep}{0.42 cm}
\begin{tabular}{cccccccc}
\hline
\hline
Refinement & Atom & Site & {\it x} & {\it y} & {\it z} & B & Occupancy \\ \hline
\multirow{4}{*}{\begin{tabular}[c]{@{}c@{}}XRD\\ x = 0.1\\ T = 280 K\\ $\chi^2$ = 0.231\\ (a)\end{tabular}} & Fe & 8a & 1/8 & 1/8 & 1/8 & 0.836(67) & 0.03750 \\
 & Co & 8a & 1/8 & 1/8 & 1/8 & 0.836(67) & 0.00417 \\
 & V & 16d & 1/2 & 1/2 & 1/2 & 0.638(53) & 0.08330 \\
 & O & 32e & 0.75350(77) & 0.75350(77) & 0.75350(77) & 3.914(146) & 0.16670 \\
\multicolumn{1}{l}{} & \multicolumn{1}{l}{} & \multicolumn{1}{l}{} & \multicolumn{1}{l}{} & \multicolumn{1}{l}{} & \multicolumn{1}{l}{} & \multicolumn{1}{l}{} & \multicolumn{1}{l}{} \\
 & \multicolumn{6}{c}{a = b = c = 8.46316(6)} &  \\ \hline
\multirow{4}{*}{\begin{tabular}[c]{@{}c@{}}XRD\\ x = 0.1\\ T = 10 K\\ $\chi^2$ = 0.386\\ (b)\end{tabular}} & Fe & 4b & 1/2 & 1/4 & 1/8 & 0.836(57) & 0.11250 \\
 & Co & 4b & 1/2 & 1/4 & 1/8 & 0.836(57) & 0.01125 \\
 & V & 8c & 1/4 & 3/4 & 1/4 & 0.286(43) & 0.25000 \\
 & O & 16h & 1/2 & 0.99279(69) & 0.25734(46) & 3.244(112) & 0.50000 \\
\multicolumn{1}{l}{} & \multicolumn{1}{l}{} & \multicolumn{1}{l}{} & \multicolumn{1}{l}{} & \multicolumn{1}{l}{} & \multicolumn{1}{l}{} & \multicolumn{1}{l}{} & \multicolumn{1}{l}{} \\
 & \multicolumn{6}{c}{a = b = 5.94631(14), c = 8.53846(22)} &  \\ \hline
\multirow{4}{*}{\begin{tabular}[c]{@{}c@{}}XRD\\ x = 0.5\\ T = 280 K\\ $\chi^2$ = 0.406\\ (c)\end{tabular}} & Fe & 8a & 1/8 & 1/8 & 1/8 & 0.630(81) & 0.02083 \\
 & Co & 8a & 1/8 & 1/8 & 1/8 & 0.630(81) & 0.02083 \\
 & V & 16d & 1/2 & 1/2 & 1/2 & 2.825(95) & 0.08330 \\
 & O & 32e & 0.73756(41) & 0.73756(41) & 0.73756(41) & 2.580(194) & 0.16670 \\
\multicolumn{1}{l}{} & \multicolumn{1}{l}{} & \multicolumn{1}{l}{} & \multicolumn{1}{l}{} & \multicolumn{1}{l}{} & \multicolumn{1}{l}{} & \multicolumn{1}{l}{} & \multicolumn{1}{l}{} \\
 & \multicolumn{6}{c}{a = b = c = 8.45129(24)} &  \\ \hline
\multirow{4}{*}{\begin{tabular}[c]{@{}c@{}}XRD\\ x = 0.5\\ T = 10 K\\ $\chi^2$ = 0.254\\ (d)\end{tabular}} & Fe & 4b & 1/2 & 1/4 & 1/8 & 0.138(105) & 0.06250 \\
 & Co & 4b & 1/2 & 1/4 & 1/8 & 0.138(105) & 0.06250 \\
 & V & 8c & 1/4 & 3/4 & 1/4 & 0.234(91) & 0.25000 \\
 & O & 16h & 1/2 & 0.98815(129) & 0.25334(96) & 1.229(196) & 0.50000 \\
\multicolumn{1}{l}{} & \multicolumn{1}{l}{} & \multicolumn{1}{l}{} & \multicolumn{1}{l}{} & \multicolumn{1}{l}{} & \multicolumn{1}{l}{} & \multicolumn{1}{l}{} & \multicolumn{1}{l}{} \\
 & \multicolumn{6}{c}{a = b = 5.95824(29), c = 8.47943(43)} & \multicolumn{1}{l}{} \\ \hline \hline
\end{tabular}
\end{table*}

\subsection{IV. DISCUSSION}

Based on the magnetic and structural data, several observations can be made. First, the transition from the cubic to the HT tetragonal ($c$ $<$ $a$) phase at $T_1$ and the transition from the HT tetragonal to the orthorhombic phase at $T_2$ appear for the $x$ = 0 and 0.05 samples but disappear for the $x$ $\geq$ 0.1 samples. This suggests that slight disorder or Co-doping on the Fe sites is sufficient to suppress both transitions, behavior which confirms that both structural phase transitions are dominated by the A site Fe$^{2+}$ ions. The transition at $T_1$ is due to the JT type compression of the FeO$_4$ tetrahedron, and the transition at $T_2$ is due to the spin-orbital interaction of the Fe$^{2+}$ ions in the magnetic ordered phase\cite{Fe4, Fe5}. Second, for the $x$ $\geq$ 0.1 samples, the paramagnetic to CF transition temperature ($T_2$) increases with increasing Co-doping. As shown in Fig. 9(a), the V-V distance decreases with increasing Co-doping. This is similar to the chemical pressure effects on Mn$_{1-x}$Co$_x$V$_2$O$_{4}$. The resistivity studies on Mn$_{1-x}$Co$_x$V$_2$O$_{4}$\cite{MnCo1} have shown that with decreasing V-V distance the system approaches the itinerant electron behavior. The DFT calculation on CoV$_2$O$_4$\cite{MnCo3} then shows that this increasing electronic itinerancy  can lessen the magnetic anisotropies and enhance the A-B site's magnetic exchange interactions to increase the CF transition temperature. This increase of $T_2$ with increasing Co-doping  in Mn$_{1-x}$Co$_x$V$_2$O$_{4}$ has been experimentally confirmed, and we believe a similar situation occurs with increasing Co-doping in Fe$_{1-x}$Co$_x$V$_2$O$_{4}$. Third, for the $x$ = 0.1 and 0.2 samples, a cubic to LT tetragonal ($c$ $>$ $a$) phase transition occurs around the paramagnetic to CF transition at $T_2$, but for the 0.3 $\leq$ $x$ $\leq$ 0.6 samples, this structural phase transition occurs at $T_6$ which is below the CF transition temperature $T_2$. For the $x$ $\geq$ 0.7 samples, no structural phase transition is observed down to 10 K. The direct change from the cubic to tetragonal ( $c$ $>$ $a$) phase shows that for the 0.1 $\leq$ $x$ $\leq$ 0.6 samples with larger doping on the Fe$^{2+}$ sites, this transition is controlled by the ferroic-orbital ordering of the V$^{3+}$ ions. The decoupling of the magnetic phase transition at $T_2$ and structural phase transition at $T_6$ for the 0.3 $\leq$ $x$ $\leq$ 0.6 samples show the competition between the orbital ordering and itinerancy of V$^{3+}$ electrons. With increasing Co-doping, the increasing electronic itinerancy leads to enhanced magnetic ordering that contrasts with the decreasing orbital ordering which is completely suppressed for the $x$ $\geq$ 0.7 samples. This is also revealed by the decreasing $c/a$ ratio (decreasing distortion, without distortion $c/a$ = 1.0 for the $x$ $\geq$ 0.7 samples) with increasing Co-doping.

Some other details of the phase diagram are: (i) the CF-NCF transition temperature $T_5$ ($\sim$ 60 K) for  0 $\leq$ x $\leq$ 0.8 is doping-independent. It jumps to 75 K ($T_4$) for $x$ = 0.9 and 1.0 samples. For the $x$ = 0 and 0.05 samples, the orthorhombic to tetragonal ($c$ $>$ $a$) structural phase transition ($T_3$) occurs simultaneously at $T_5$. However, for the 0.1 $\leq$ x $\leq$ 0.7 samples, there is no structural phase transition at $T_5$. This indicates that in this regime, the $T_5$ (NCF magnetic ordering) is controlled only by the V$^{3+}$ ions. Then for the $x$ = 0.9 and 1.0 samples, the enhanced magnetic exchange isotropy due to the stronger electronic itinerancy stabilizes the CF-NCF transition at 75 K\cite{MnCo3} which has been demonstrated by the DFT calculations on CoV$_2$O$_4$. The derivative of the susceptibility of the $x$ = 0.8 sample shows two features for the NCF ordering: a  broad peak at $T_5$ = 60 K similar to that of the 0.1 $\leq$ x $\leq$ 0.7 samples and a sharp peak at 40 K similar to that of the $x$ $\leq$ 0.9 samples. This suggests that the $x$ = 0.8 sample is on the boundary for the competitions between the orbital ordering of the localized V$^{3+}$ spins and the enhanced exchange isotropy due to the itinerancy. The former stabilizes the NCF ordering at $T_5$ while the latter stabilizes the NCF ordering at 40 K and then improves it to $T_4$ =75 K for the $x$ $\geq$ 0.9 samples; (ii) for the $x$ = 0.1 and 0.2 samples, there is no structural phase transition at $T_3$. They instead exhibit a subtle structural distortion with decreased $c/a$ ratio. Moreover, this particular $T_3$ is below the CF-NCF transition temperature, $T_5$. In the Fe$_{1-x}$Mn$_x$V$_2$O$_{4}$ system, a similar decreased $c/a$ ratio has also been observed at the CF-NCF transition temperature which indicates this subtle structural distortion is due to the spin-lattice coupling of the V spin-canting process. Despite the decoupling of $T_3$ and $T_5$ here, a similar situation may occur around $T_3$ for the $x$ = 0.1 and 0.2 samples. 

We compare the phase diagram between Fe$_{1-x}$Co$_x$V$_2$O$_4$ and Fe$_{1-x}$Mn$_x$V$_2$O$_4$. The similarity is that in both systems, the HT tetragonal and orthorhombic phases disappear quickly with small doping. This again confirms both phases are due to the presence of the Fe$^{2+}$ ions on the A sites. The main difference is that in  Fe$_{1-x}$Mn$_x$V$_2$O$_4$, the paramagnetic to CF transition is always accompanied with the cubic to tetragonal phase transition for the $x$ $\leq$ 0.6 samples, and the CF to NCF transition is always accompanied with another type of cubic to tetragonal phase transition for the $x$ $\geq$ 0.7 samples. In other words,  the spin ordering and structural phase transition are  always strongly coupled for Fe$_{1-x}$Mn$_x$V$_2$O$_4$. However, in Fe$_{1-x}$Co$_x$V$_2$O$_4$ these two transitions are decoupled with the structural phase transition occurring below the paramagnetic to CF transition. Meanwhile, in the  Mn$_{1-x}$Co$_x$V$_2$O$_4$ phase diagram, the CF-NCF transition is decoupled from the cubic-tetragonal structural phase transition. For MnV$_2$O$_4$, both transitions occur at the same temperature, but with increasing Co-doping in Mn$_{1-x}$Co$_x$V$_2$O$_4$ the CF-NCF  transition occurs at higher temperatures and is followed by the structural phase transition at lower temperatures. This is similar to the separation between $T_3$ and $T_5$ for the $x$ = 0.1 and 0.2 samples in  Fe$_{1-x}$Co$_x$V$_2$O$_4$. Therefore, one general behavior for Co-doping systems seems to be the separation of the magnetic and structural phase transitions. This separation should be due to the induced competition between the orbital ordering and electronic itinerancy. With increasing Co-doping, the increased electronic itinerancy tends to enhance the A-B magnetic interaction and magnetic exchange isotropy ( to increase the CF and NCF transition temperatures) and suppress the orbital ordering (the structural phase transition temperature).

\subsection{V. CONCLUSION}
In summary, the single crystals of Fe$_{1-x}$Co$_x$V$_2$O$_{4}$ were studied by specific heat, susceptibility, elastic neutron scattering, and XRD measurements. The main findings are with increasing Co-doping: (i) the HT tetragonal and orthorhombic phases disappear quickly due to the small disorder on the Fe$^{2+}$ sites. This confirms these two phases are due to the JT type distortion and spin-orbital coupling of the Fe$^{2+}$ ions; (ii) the increased electronic intinerancy results in enhanced magnetic ordering but suppressed orbital ordering. The consequence is a complex magnetic and structural phase diagram with decoupled magnetic and structural phase transition boundaries.

\subsection{ACKNOWLEDGEMENTS}
\begin{acknowledgments}
R. S., Z.L.D., and H.D.Z. thank the support from NSF-DMR through Award DMR-1350002. The research at HFIR/ORNL, were sponsored by the Scientific User Facilities Division (J.M., H.B.C., T.H.,  M.M.,), Office of Basic Energy Sciences, US Department of Energy.
\end{acknowledgments}


\begin{thebibliography}{99}
\bibitem{AV2O4review} S.~-H.~Lee, H.~Takagi, D.~Louca, M.~Matsuda, S.~Ji, H.~Ueda, Y.~Ueda, T.~Katsufuji, J.~-H.~Chung, S.~Park, S.~-W.~Cheong, and C.~Broholm, J. Phys. Soc. Jpn. {\bf 79}, 011004  (2010).
\bibitem{Cd1}M.~Onoda and J.~Hasegawa, J. Phys.: Condensed Matter {\bf 15}, L95  (2003).
\bibitem{Cd2}G.~Giovannetti, A.~Stroppa, S.~Picozzi, D.~Baldomir, V.~Pardo, S.~Blanco-Canosa, F.~Rivadulla, S.~Jodlauk, D.~Niermann, J.~Rohrkamp, T.~Lorenz, S.~Streltsov, D.~I.~Khomskii, and J.~Hemberger, Phys. Rev. B {\bf 83}, 060402(R) (2011).
\bibitem{Cd3}A.~N.~Vasiliev, M.~M.~Markina, M.~Isobe, Y.~Ueda, J. Magn. Magn. Mater. {\bf 300}, e375 (2006).
\bibitem{Mg1}H.~Mamiya, M.~Onoda, T.~Furubayashi, J.~Tang, and I.~Nakatani, J. Appl. Phys. {\bf 81}, 5289  (1997).
\bibitem{Mg2}E.~M.~Wheeler, B.~Lake, A.~T.~M.~N.~Islam, M.~Reehuis, P.~Steffens, T.~Guidi, and A.~H.~Hill, Phys. Rev. B {\bf 82}, 140406(R) (2010).
\bibitem{Mg3}A.~T.~M.~N.~Islam, E.~M.~Wheeler, M.~Reehuis, K.~Siemensmeyer, M.~Tovar, B.~Klemke, K.~Kiefer,  A.~H.~Hill, and B.~Lake, Phys. Rev. B {\bf 85}, 024203 (2012).
\bibitem{Mg4}S. Niitaka, H.~Ohsumi, K.~Sugimoto, S.~Lee, Y.~Oshima, K.~Kato, D.~Hashizume, T.~Arima, M.~Takata, and H.~Takagi, Phys. Rev. Lett. {\bf 111}, 267201 (2013).
\bibitem{Mg5}E.~D.~Mun, G.~W.~Chern, V.~Pardo, F.~Rivadulla, R.~Sinclair, H.~D.~Zhou, V.~S.~Zapf, and C.~D.~Batista, Phys. Rev. Lett. {\bf 112}, 017207 (2014).
\bibitem{Zn1}M.~Reehuis, A.~Krimmel, N.~B\"{u}ttgen, A.~Loidl, and A.~Prokofiev, Eur. Phys. J. B {\bf 35}, 311  (2003).
\bibitem{Zn2}S.~-H.~Lee, D.~Louca, H.~Ueda, S.~Park, T.~J.~Sato, M.~Isobe, Y.~Ueda, S.~Rosenkranz, P.~Zschack, J.~\'{I}\~{n}iguez, Y.~Qiu, and R.~Osborn, Phys. Rev. Lett. {\bf 93}, 156407 (2004).
\bibitem{Mn1}K.~Adachi, T.~Suzuki, K.~Kato, K.~Osaka, M.~ Takata, and T.~Katsufuji, Phys. Rev. Lett. {\bf 95},197202  (2005).
\bibitem{Mn2}S.~-H.~Baek, N.~J.~Curro, K.~-Y.~Choi, A.~P.~Reyes, P.~L.~Kuhns, H.~D.~Zhou, and C.~R.~Wiebe, Phys. Rev. B {\bf 80}, 140406(R), (2009).
\bibitem{Mn3}K.~Myung-Whun, J.~S.~Kim, T.~Katsufuji, and R. K. Kremer, Phys. Rev. B {\bf83}, 024403 (2011).
\bibitem{Mn4}K.~Takubo, R.~Kubota, T.~Suzuki, T.~Kanzaki, S.~Miyahara, N.~Furukawa, and T.~Katsufuji, Phys. Rev. B {\bf 84}, 094406 (2011).
\bibitem{Mn5}V.~O.~Garlea, R.~Jin, D.~Mandrus, B.~Roessli, Q.~Huang, M.~Miller, A.~J.~Schultz, and S.~E.~Nagler, Phys. Rev. Lett. {\bf 100}, 066404 (2008).
\bibitem{Mn6}H.~D.~Zhou, J.~Lu, and C.~R.~Wiebe, Phys. Rev. B {\bf76}, 174403 (2007).
\bibitem{Mn7}S.~L.~Gleason, T.~Byrum, Y.~Gim, A.~Thaler, P.~Abbamonte, G.~J.~MacDougall, L.~W.~Martin, H.~D.~Zhou, and S.~L.~Cooper, Phys. Rev. B {\bf 89} 134402 (2014).
\bibitem{Co1} A.~Kismarahardja, J.~S.~Brooks, A.~Kiswandhi, K.~Matsubayashi, R.~Yamanaka, Y.~Uwatoko, J.~Whalen, T.~Siegrist, and H.~D.~Zhou, Phys. Rev. Lett. {\bf 106}, 056602 (2011).
\bibitem{Co2} Y.~J.~Huang, Z.~R.~yang, and Y.~H.~Zhang, J. Phys.: Condens. Matt. {\bf 24} 056003 (2012).
\bibitem{Co3} R.~Kaur, T.~Maitra, and T.~Nautiyal, J. Phys.: Condens. Matt. {\bf 26} 045505 (2014).
\bibitem{Goodenough} J.~B.~Goodenough, in {\em Metallic Oxides}, edited by H. Reiss, Progress in Solid State Chemistry Vol. 5 (Pergamon, New York, 1972).
\bibitem{Fran2} V.~Pardo, S.~Blanco-Canosa, F.~Rivadulla, D.~I.~Khomskii, D.~Baldomir, Hua Wu, and J.~Rivas, Phys. Rev. Lett. {\bf 101}, 256403 (2008).
\bibitem{Rogers1} D.~B.~Rogers, R.~J.~Arnott, A.~Wold, and J.~B.~Goodenough, J. Phys. Chem. Solids {\bf 24}, 347 (1963).
\bibitem{Rogers2} D.~B.~Rogers, J.~B.~Goodenough, and A.~Wold, J. Appl. Phys. {\bf 35}, 1069 (1964).
\bibitem{Fran1} S.~Blanco-Canosa, F.~Rivadulla, V.~Pardo, D.~Baldomir, J.~-S.~Zhou, M.~Garc\'{i}a-Hern\'{a}ndez, M.~A.~L\'{o}pez-Quintela, J.~Rivas, and J.~B.~Goodenough, Phys. Rev. Lett. {\bf 99}, 187201 (2007).
\bibitem{Fe1}M.~Tanaka, T.~Tokoro, and Y.~Aiyama, J. Phys. Soc. Jpn. {\bf 21}, 262 (1966).
\bibitem{Fe2}T.~Katsufuji, T.~Suzuki, H.~Takei, M.~Shingu, K.~Kato, K.~Osaka, M.~Takata, H.~Sagayama, and T.~Arima, J. Phys. Soc. Jpn.{\bf 77}, 053708  (2008).
\bibitem{Fe3} Q.~Zhang, K.~Singh, F.~Guillou, C.~Simon, Y.~Breard, V.~Caignaert, and V.~Hardy, Phys. Rev. B {\bf 85} 054405 (2012).
\bibitem{Fe4} G.~J.~MacDougall, V.~O.~Garlea, A.~A.~Aczel, H.~D.~Zhou, and S.~E.~Nagler, Phys. Rev. B {\bf 86}, 060414 (2012).
\bibitem{Fe5} Y.~Nii, H.~Sagayama, T.~Arima, S.~Aoyagi, R.~Sakai, S.~Maki, E.~Nishibori, H.~Sawa, K.~Sugimoto, H.~Ohsumi, and M.~Takata, Phys. Rev. B {\bf 86} 125142 (2012).
\bibitem{Fe6} Y.~Ishitsuka, T.~Ishikawa, R.~Koborinai, T.~Omura, and T.~Katsufuji, Phys. Rev. B {\bf 90} 224411 (2014).
\bibitem{MnCo1} A.~Kiswandhi, J.~S.Brooks, J.~Lu, J.~Whalen, T.~Siegrist, and H.~D.~Zhou, Phys. Rev. B {\bf 84} 205138 (2011).
\bibitem{MnCo2} A.~Kiswandhi, J.~Ma, J.~S.Brooks, and H.~D.~Zhou, Phys. Rev. B {\bf 90} 155132 (2014).
\bibitem{MnCo3} J.~Ma, J.~H.~Lee, S.~E.~Hahn, Tao Hong, H.~B.~Cao, A.~A.~Aczel, Z.~L.~Dun, M.~B.~Stone, W.~Tian, Y.~Qiu, J.~R.~D.~Copley, H.~D.~Zhou, R.~S.~Fishman, and M.~Matsuda, Phys. Rev. B {\bf 91} 020407(R) (2015).
\bibitem{FeMn1} D.~Choudhury, T.~Suzuki, D.~Okuyama, D.~Morikawa, K.~Kato, M.~Takata, K.~Kobayashi, R.~Kumai, H.~Nakao, Y.~Murakami, M.~Bremholm, B.~B.~Iversen, T.~Arima, Y.~Tokura, and Y.~Taguchi, Phys. Rev. B {\bf 89} 104427 (2014).
\bibitem{FeMn2} S.~Kawaguchi, H.~Ishibashi, J.~Kim, and Y.~Kubota, J. Phys.: Condens. Matt. {\bf 26} 346001 (2014).

\end{thebibliography}
\end{document}